 \documentclass[manuscript]{aastex}

\def\dd{{\bf {d}}}

\def\nn{{\bf {n}}}

\def\xx{{\bf {x}}}

\def\BB{{\bf {B}}}

\def\JJ{{\bf {J}}}

\def\xhat{{\bf {\hat x}}}
\def\yhat{{\bf {\hat y}}}
\def\zhat{{\bf {\hat z}}}

\shorttitle{Lagrangian relaxation schemes for calculating force-free fields}
\shortauthors{Pontin et al.}

\begin{document}


\title{Lagrangian relaxation schemes for calculating force-free magnetic fields, and their limitations}


\author{D. I. Pontin, G. Hornig and A. L. Wilmot-Smith}
\affil{Division of Mathematics, University of Dundee, Dundee, UK}

\and

\author{I. J. D. Craig}
\affil{Mathematics Department, University of Waikato, Hamilton, New Zealand}




\begin{abstract}
Force-free magnetic fields are important in many astrophysical settings. Determining the properties of such force-free fields -- especially smoothness and stability properties -- is crucial to understanding many key phenomena in astrophysical plasmas, for example energy release processes that heat the plasma and lead to dynamic or explosive events. 
{Here we report on a serious limitation on the computation of force-free fields that has the potential to invalidate the results produced by numerical force-free field solvers even for cases in which they appear to converge 
{(at fixed grid resolution)} to an equilibrium magnetic field. In the present   work we discuss this problem within the context of a Lagrangian relaxation scheme that conserves magnetic flux and $\nabla\cdot\BB$ identically.}
Error estimates are introduced  to assess the quality of the calculated equilibrium.  We go on to present an algorithm, based  on re-writing the $curl$ operation via Stokes' theorem,  for calculating the current which holds great promise for improving dramatically the accuracy of the {Lagrangian} relaxation procedure.
\end{abstract}



\section{Introduction}
Force-free magnetic fields, $\BB$, satisfy $\JJ\times\BB={\bf 0}$, or equivalently
\begin{equation}
(\nabla \times {\bf B})\times \BB = {\bf 0}.
\end{equation}
Calculation of such force-free fields is of importance in many astrophysical settings, for example accretion disks around various objects  \citep[e.g.][]{frank2002, uzdensky2002}, neutron stars \citep{mckinney2006}, pulsars \citep{mestel1973}, magnetic clouds \citep{burlaga1988}, and solar and stellar coronae \citep[e.g.][]{anzer1968}.

A particular application in solar physics is the controversial  `topological dissipation' model proposed by \cite{parker1972}. The assertion of this model is that if an equilibrium magnetic field is perturbed by arbitrary motions at a line-tied boundary, then the subsequent field cannot relax to a smooth force-free equilibrium. Rather, the equilibrium must contain tangential discontinuities -- corresponding to current sheets. Doubt has been cast upon the  model however, as a number of authors have demonstrated the existence of smooth solutions in the scenario posed \citep{vanballegooijen1985, zweibel1987, longcope1994, craigsneyd2005}. The question as to  whether current sheets form spontaneously in the coronal magnetic field is key to understanding the so-called coronal heating problem.
This is just one example which demonstrates that determining both the structure and stability of force-free magnetic fields is of fundamental importance.

There are different approaches that one may take when searching for force-free magnetic fields. One method, often used when modelling the solar corona, is to solve a boundary value problem (\cite{amari1997}, and see \cite{schrijver2006} for a comparison of numerical schemes). The force-free field is reconstructed from boundary data, provided for example by a vector magnetogram.  An alternative approach is to begin with an initial magnetic field that is not force-free and to perform a relaxation procedure. This is the natural approach if one wants to investigate the properties of particular magnetic topologies. As long as the relaxation procedure can be guaranteed to be ideal, then the topology will be conserved during the relaxation.

One powerful computational approach for 
investigating the properties of force-free fields is to employ an ideal Lagrangian relaxation scheme. Such schemes exploit the property that under ideal MHD the vector ${\bf B}/\rho$ evolves according to the equation
\begin{equation}
\frac{D}{Dt} \frac{{\bf B}}{\rho} = \left( \frac{{\bf B}}{\rho} \cdot \nabla \right) {\bf v}
\end{equation}
where $D/Dt$ is the material derivative, $\rho$ the plasma density and ${\bf v}$ the plasma velocity. This is of exactly the same form as the evolution equation of a line element ${\bf \delta x}$ in a flow (see, e.g.~\cite{moffatt1978}), and thus a Lagrangian description facilitates a relaxation that is, by construction, ideal. 
{These schemes can
      be used to investigate the structure and (ideal MHD) stability of force-free
      fields.  The latter is guaranteed by the iterative convergence 
      of the scheme provided that the resolution is sufficient.}
The primary variables that the numerical scheme dynamically updates are the locations of the mesh points, with the quantities ${\bf B}$ and ${\bf J}$ being calculated via matrix products involving the initial magnetic field and derivatives of the mapping that describes the mesh deformation. 
An artificial frictional term is included in the equation of motion (see also \cite{chodura1981}) which guarantees a monotonic decrease of the energy.
Two implementations of this method are described in \cite{craig1986} and \cite{longbottom1998}. The method has been used extensively to investigate the stability and equilibrium properties of various different magnetic configurations, such as the kink instability of magnetic flux tubes \citep{craig1990}, line-tied collapse of 2D and 3D magnetic null points \citep{craiglitvinenko2005, pontincraig2005} and the Parker problem \citep{longbottom1998, craigsneyd2005}.

In the following section we describe a test problem that illustrates
{one major difficulty in the computation of force-free fields, in the context of the Lagrangian relaxation scheme outlined above.}
In Section \ref{numsec} we present two possible extensions of the numerical scheme. In Section \ref{ressec} we describe our results, and in Section \ref{concsec} we present our conclusions.

\section{The problem}\label{motsec}
\subsection{Outline of the problem}
In a numerical relaxation experiment using braided initial fields \citep{wilmotsmith2009} we came across  an inconsistency of the resulting numerical force-free state, which is best explained with the help of the following example. Consider a magnetic field obtained from the homogenous field by a simple twisting deformation as shown in Fig.~\ref{twoblobfig}(a).  Obviously an ideal relaxation towards a force-free state must end again in a homogenous state ($\JJ={\bf 0}$). During this process the Lagrangian  relaxation leads to a deformation of the initial computational mesh which exactly cancels the initial deformation applied to the homogenous field. This is a well defined setup in which we know exactly the initial and final states. 
We now employ the implicit (ADI) relaxation scheme detailed by \cite{craig1986} to relax our twisted field to a force-free equilibrium.
The magnetic field is line-tied on all boundaries (${\bf B}\cdot{\hat {\bf n}}=0$ on $x$ and $y$ boundaries). The ${\bf J}\times{\bf B}$ force as calculated by the numerical scheme decreases monotonically to an arbitrarily small value (e.g.~$10^{-6}$--$10^{-8}$), giving the appearance that the scheme converges {(in an iterative sense)} to a force-free equilibrium (to any desired accuracy, down to machine precision). 
However, when plotting $\alpha$, the force-free proportionallity  factor, along a field line it shows variations which are by orders of magnitude higher than would be expected from $|\JJ\times\BB| < 10^{-8}$. It is this inconsistency that we investigate in what follows.
{As we will discuss the convergence of the numerical scheme in what follows, it is worth emphasising here the distinction between iterative 
convergence (at fixed resolution $N$) and real convergence, i.e.~convergence towards a `correct' solution as the resolution $N$ becomes sufficiently large.
}

\subsection{Analysis}
\begin{figure}[]
\centering
(a)\includegraphics[scale=0.5]{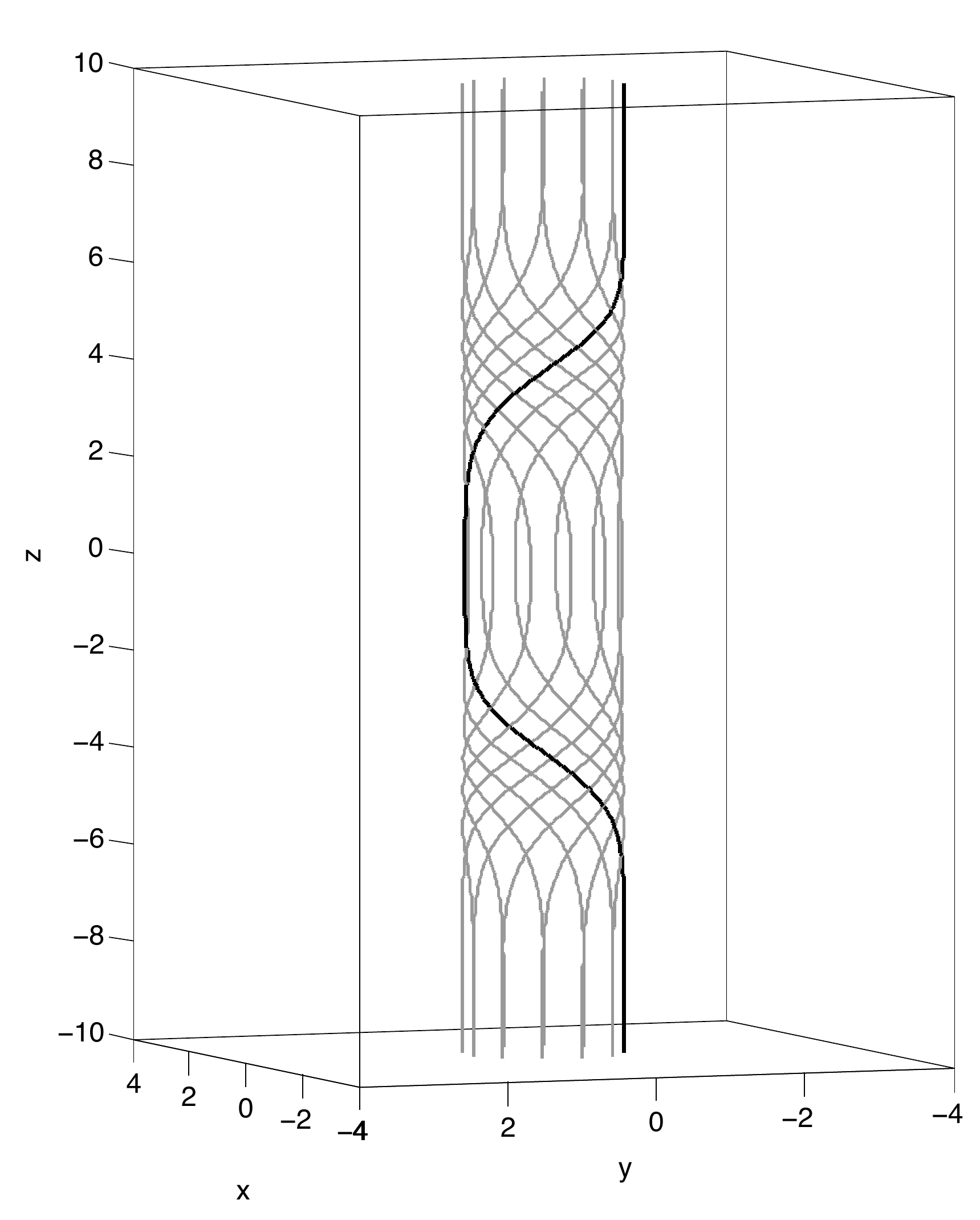}
(b)\includegraphics[scale=0.5]{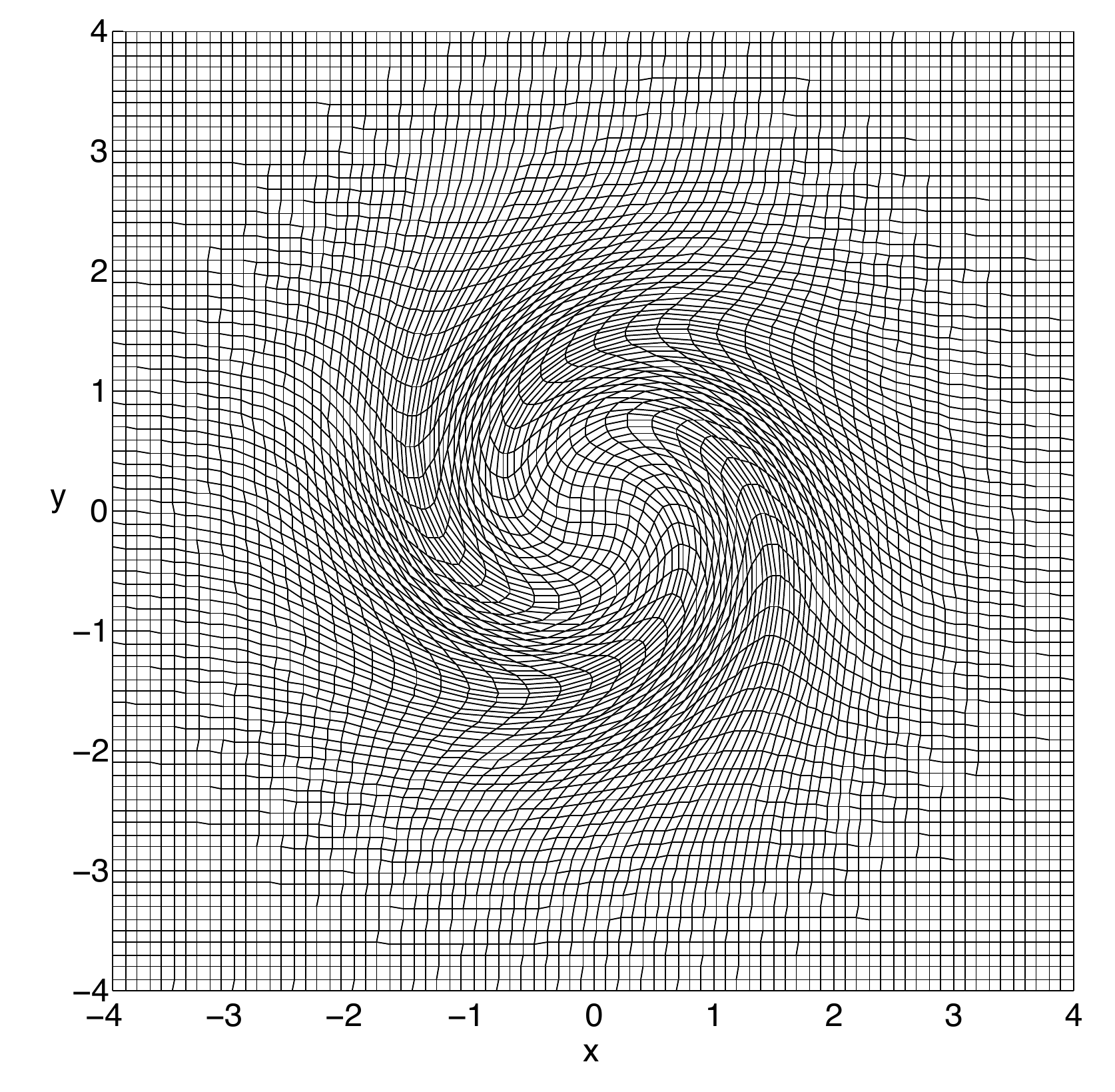}
\caption{(a) Sample field lines for the field $T2$, given by Eq.~(\ref{t2}). (b) Mesh in the $z=0$ plane for the test problem with artificially imposed deformation, with $\psi=\pi$ and resolution $81^3$.}
\label{twoblobfig}
\end{figure}

In order to investigate the source of the inconsistency described in the previous section, we consider the test problem outlined there. 
Specifically, we begin with an initially uniform magnetic field ($\BB=b_0\zhat$), and superimpose two regions of toroidal field, centred on the $z$-axis at $\pm z_0$, with exactly the same functional form, but of opposite signs:
\begin{equation}\label{t2}
\mathbf{B} = b_{0} \hat{\bf{z}} + \sum_{i=1}^{2} 
\frac{2 b_{0} \phi_i}{\pi a_r}  \exp \left(
- \frac{ \scriptstyle x^2+y^2}{ \scriptstyle a_r^{2}} 
- \frac{ \scriptstyle \left(z-L_i\right)^{2}}{\scriptstyle a_z^{2}}  \right)
\left( - y \xhat + x \yhat  \right),
\end{equation}
with $b_0=1$, $a_r=\sqrt{2}$, $a_z=2$ and $\phi_1=\pi$, $\phi_2=-\pi$, $L_1=-4$, $L_2=4$. We refer to this field in the following as $T2$. The field $T2$ (see Fig.~\ref{twoblobfig}(a)) is constructed such that  
the two regions of twisted field, which are of opposite sign, should exactly cancel one another under an ideal relaxation, approaching the uniform field (with ${\bf J}={\bf 0}$) as the equilibrium. Note that $|\phi|$ is the maximum turning angle of field lines around the $z$-axis.

One of the great advantages of an ideal Lagrangian relaxation is that it is possible to extract the paths of the magnetic field lines of the final state if one knows them in the initial state, simply by interpolating over the mesh displacement. Calculating the field lines in this way, no error is accumulated by integrating along ${\bf B}$. Given knowledge of the field line paths, one can test the quality of the force-free approximation by plotting $\alpha$ along field lines. For a force-free field
\begin{equation}\label{fff}
\nabla \times {\bf B} = \alpha {\bf B},
\end{equation}
and $\alpha$ should be constant along field lines since taking the divergence of the above yields
\begin{equation}\label{alphaalongB}
{\bf B}\cdot \nabla \alpha=0.
\end{equation}

We begin by defining the variable $\alpha^{*}$, motivated by Eq.~(\ref{fff}), as 
\begin{equation}\label{alphastar}
\alpha^* = \frac{ J_{\|} }{|{\bf B}|}.
\end{equation}
We find that for the magnetic field calculated by the relaxation scheme, the value of $\alpha^*$ changes dramatically along field lines.
Of course the relaxation gives a magnetic field for which ${\bf J}\times {\bf B}$ is not identically zero. So for a given value of ${\bf J}\times{\bf B}$, what is the maximum possible variation in $\alpha^*$ along a field line?

Consider
\begin{displaymath}
\nabla \cdot {\bf J} = \nabla \cdot (J_{\|} {\hat {\bf B}}) + \nabla \cdot {\bf J}_{\perp} = \delta,
\end{displaymath}
say, where $\delta$ is representative of the error in calculating $\JJ$.
Using Eq.~(\ref{alphastar}) to replace $J_{\|}$ gives
\begin{equation}
{\bf B} \cdot \nabla \alpha^* = - \nabla \cdot {\bf J}_{\perp} + \delta.
\end{equation}
or
\begin{equation}\label{eq8}
\frac{d\alpha^*}{dl} = - \frac{\nabla \cdot {\bf J}_{\perp}}{|{\bf B}|} + \frac{\delta}{|\BB|}.
\end{equation}
where $l$ is a parameter along a magnetic field line with units of length. Now  suppose that $|{\bf J}\times {\bf B}|/|\BB|^2 < \epsilon$ within our domain. This implies that $|{\bf J}_{\perp}| < \epsilon \, |{\bf B}|$, so that
\begin{displaymath}
|\nabla \cdot {\bf J}_{\perp}| < \frac{\epsilon \,|{\bf B}|}{ d},
\end{displaymath}
where $d$ is the length scale of variations perpendicular to the magnetic field. Then from Eq.~(\ref{eq8})
\begin{equation}\label{fffquality}
\left| \frac{d \alpha^*}{dl}\right| < \frac{\epsilon}{d} + \frac{|\delta|}{|\BB|}.
\end{equation}

Returning to our relaxation results, we have for example $\epsilon=10^{-6}$, with $|{\bf B}|\approx 1$,  $d \approx \sqrt{2}$. 
However, we find that 
$|d\alpha^*/dl |_{max} \approx 0.02 $. The discrepancy between this figure and the value of $\epsilon$ must come from the final term in Eq.~(\ref{fffquality}). This has been checked by interpolating the data onto a rectangular mesh and approximating $\nabla\cdot\JJ$ using standard finite differences. We find $\nabla\cdot\JJ \sim O(10^{-2})$, and it therefore appears that the residual currents parallel to $\BB$ are not relaxed because $\nabla \cdot {\bf J} \neq 0$. {As demonstrated below, this error does however decrease as the resolution is increased (see Tables \ref{errortab}--\ref{tab3}).}

It turns out that the appearance of the errors is related to the way in which ${\bf J}$ is calculated within the scheme, via a combination of 1st and 2nd derivatives of the deformation matrix. These derivatives are calculated via  finite differences in the numerical scheme, and it is here that these {discretisation} errors arise. This is demonstrated below.

\subsection{Accuracy test: artificially imposed deformation}
To ascertain the source of the errors, we take our initial state $T2$ and {\it instead of performing the relaxation procedure}, we artificially apply a deformation to the mesh which we can write down as a closed form expression, and moreover for which we can obtain the derivatives of the mesh displacement, and thus the resultant ${\bf B}$ and ${\bf J}$ fields, as closed form expressions. Motivated by the results of the relaxation, we impose a similar rotational distortion of the mesh which acts to `untwist' the field, via the transformation
\begin{equation}
(x,y,z) \longrightarrow (x \cos \theta -y \sin \theta ~,~ y\cos\theta +x\sin\theta ~,~ z)
\end{equation}
where
\begin{equation}\label{analytangle}
\theta = \psi \exp\left( -\frac{x^2+y^2}{4} -\frac{z^2}{16} \right),
\end{equation}
$\psi$ constant. 
We now apply this transformation to an initially rectangular mesh on which $\BB$ is given by $T2$,  and compare the numerical and exact values for each entry in the mesh deformation Jacobian, and each component of ${\bf B}$ and ${\bf J}$. Results are shown for three different values of the parameter $\psi$ in Table \ref{errortab}.

\begin{table}
\centering
\begin{tabular}{c || c | c | c}
$N$ & ~$\psi=\pi$~ & ~$\psi=3\pi/4$~ & ~$\psi=\pi/2$~ \\ \cline{2-4} \hline 
21 & 29.5 &  15.3 & 6.46  \\ \cline{2-4}
	& 670  & 159 & 37.8  \\ \hline &&& \\[-1.4em] \hline
41 & 7.22 &  3.93 &  1.72  \\ \cline{2-4}
	& 144 &  42.4 & 10.6  \\ \hline &&& \\[-1.4em] \hline
61 & 3.30 &  1.80 &  0.793  \\ \cline{2-4}
	& 64.7 &  19.9 & 4.99  \\ \hline &&& \\[-1.4em] \hline
81 & 1.86  & 1.02 & 0.451  \\\cline{2-4}
	& 37.0 &  11.3 & 2.83 
\end{tabular}
\caption{Errors in ${\bf B}$ and ${\bf J}$ for deformations with $\psi=\pi, \, 3\pi/4, \, \pi/2$ (in Eq.~(\ref{analytangle})) using 2nd-order finite differences. $N^3$ is the mesh resolution. In each case the upper number shows the maximum relative percentage error in the domain over all components of ${\bf B}$, i.e. $100\times |B_i - B_i^a |_{max} / |B_i^a |_{max}$, where ${\bf B}^a$ is the exact value. The lower number is $100\times |J_i - J_i^a |_{max} / |J_i^a |_{max}$.}
\label{errortab}
\end{table}

It is clear that there are large errors in the current calculated by the numerical scheme. While errors in ${\bf B}$ and in each individual term in the mesh distortion Jacobian are much smaller, it turns out that the combination in which they are multiplied, summed, and divided to calculate ${\bf J}$ incurs large errors. The calculation has been meticulously checked such that we are certain that the error appears not due to mathematical or coding error, but rather due to an accumulation of numerical {truncation errors} in the process of calculating ${\bf J}$ \citep[via Eq.~(2.10) in][]{craig1986}. Note that the errors increase as the mesh distortion ($\psi$) increases, and decrease with resolution ($N$).

\section{More sophisticated numerical schemes}\label{numsec}
\subsection{Higher-order derivatives}
Clearly the accuracy of the force-free approximation is impaired by the accuracy of $curl$ operation performed in the numerical scheme.
One way to increase the accuracy of spatial derivatives  could be to use higher-order finite difference expressions. Existing versions of the scheme use conventional second-order centred-differences involving two nearest-neighbour (n.n.) values. If we expect smooth solutions (without grid-scale features) then increasing to fourth-order finite difference expressions (using 4 n.n.) is expected to increase the accuracy.

{Recall that in the frictional Lagrangian method fluid displacements 
are determined from} 
{an equation of the form 
$$
\frac{{\partial x_i}}{\partial t} = A_{i \alpha \beta \gamma} \, \, x_{\alpha, \beta \gamma} + C_i,
$$
where $A$ and $C$ are prescribed tensor and vector functions and 
summation over repeated (Greek) indices is assumed.   Note that  
partial differentiation with respect to the background 
Cartesian cordinates $(X_1, X_2, X_3) $ is indicated using the comma notation 
(i.e. $ {\partial^2 f}/ \partial \beta \partial \gamma
= f_{,\beta\gamma} $).    }

{ The important point for us is that the method involves two spatial 
derivatives of the Lagrangian variables  $ x_\alpha $.  This reflects
the fact that the Lorentz force is computed using first and second order
derivatives of the Lagrangian mesh.   Now ``diagonal
derivatives'' such as $ x_{i, j j} $  are relatively 
easy to compute using finite differences:  they involve the point itself
and two/four nearest neighbours depending on whether the scheme is 
second or fourth order.  It is these derivatives that are handled
implicitly (via tri-diagnonal and penta-diagonal implementations of 
the ADI method) to provide,  formally at least,  the unconditional 
stability of the numerical scheme.  Note that computation
of the mixed derivatives can be more complicated:  terms 
such as $ x_{i, j k} $ 
involve sixteen terms in the fourth order scheme,  as opposed to just 
four when the method is second order.   However,  irrespective of the
formal accuracy,   perhaps the main drawback of the scheme is
that,  unlike  $ \nabla \cdot {\bf B} $,   the numerical 
evaluation of $ \nabla \cdot {\bf J} $ is not 
guaranteed to vanish identically.   
{Thus the accuracy  of the relaxed solution can be compromised by the presence of   rogue currents especially in weak field regions where the mesh is highly distorted.}
The examples presented below show
explicitly that this can restrict the convergence of the solution 
with resolution $ N $.    }

\subsection{A routine based on Stokes' theorem}\label{stokessec}
Here we present an algorithm for calculating the $curl$ of a vector field (say ${\bf J}=\nabla \times {\bf B}$) on a non-uniform mesh. This algorithm gives promising results, as discussed below, and is based on re-writing the curl operation, via Stokes' theorem, as a line integral:
\begin{equation}\label{stokes}
\oint_C {\bf B}\cdot {\bf dr} = \int_{U(C)} \JJ \cdot \hat{{\bf n}} ~dS = \int_{U(C)} {\bf J}_n ~dS
\end{equation}
where the surface $U(C)$, with unit normal ${\hat \nn}$, has the closed curve $C$ as its boundary. 
The idea is similar to that of \cite{hyman1997} who have applied such so-called `mimetic' numerical methods to solving Maxwell's equations \citep{hyman1999}. Our algorithm differs in some ways from theirs.

Eq.~(\ref{stokes}) can be discretised as follows. Suppose that we want to calculate ${\bf J}$ at the mesh point $X_{i,j,k}$. 
\begin{figure}[t]
\centering
\includegraphics[scale=0.7, angle=270]{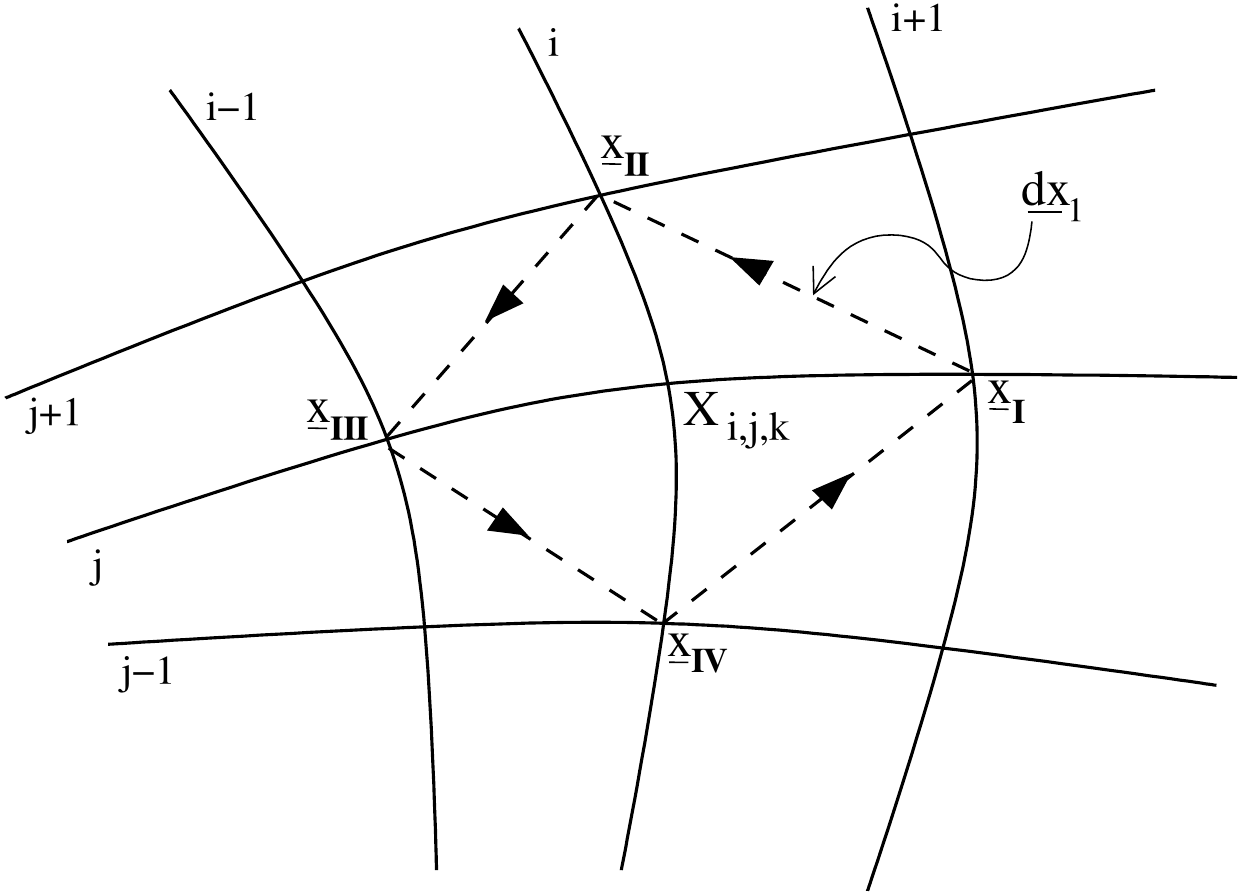}
\caption{Notation used for calculation of $\JJ$ via the Stokes-based routine.}
\label{meshpic}
\end{figure}
There are three mesh surfaces that intersect at this point. The first is the $i$th mesh level in the first index direction. Consider the circuit in this surface shown in Fig.~\ref{meshpic}, and let the nearest neighbour points to $X_{i,j,k}$ be denoted $\xx_I$, $\xx_{II}$, $\xx_{III}$, $\xx_{IV}$. Then, defining $\dd\xx_1=\xx_{II}-\xx_{I}$, $\dd\xx_2=\xx_{III}-\xx_{II}$, 
etc.~and $\BB_1=(\BB(\xx_I)+\BB(\xx_{II}))/2$, $\BB_2=(\BB(\xx_{II})+\BB(\xx_{III}))/2$ etc., we can approximate the left hand side of Eq.~(\ref{stokes}) by
\begin{equation}
\mathcal{I} = \BB_1\cdot \dd\xx_1 + \BB_2\cdot \dd\xx_2 + \BB_3\cdot \dd\xx_3 + \BB_4\cdot \dd\xx_4.
\end{equation}
Furthermore, the area of the enclosed quadrilateral is 
\begin{equation}\label{stokesarea}
A= \frac{1}{4}\, |{\bf dx}_1 \times {\bf dx}_2| + \frac{1}{4}\, |{\bf dx}_2 \times {\bf dx}_3|
+ \frac{1}{4}\, |{\bf dx}_3 \times {\bf dx}_4| + \frac{1}{4}\, |{\bf dx}_4 \times {\bf dx}_1|
\end{equation}
We can now define the direction perpendicular to this mesh surface as
\begin{equation}
\nn^{(1)} = \frac{1}{4}\left( 
	\frac{{\bf dx}_1 \times {\bf dx}_2}{|{\bf dx}_1 \times {\bf dx}_2|} + 
	\frac{{\bf dx}_2 \times {\bf dx}_3}{|{\bf dx}_2 \times {\bf dx}_3|} +
	\frac{{\bf dx}_3 \times {\bf dx}_4}{|{\bf dx}_3 \times {\bf dx}_4|} +
	\frac{{\bf dx}_4 \times {\bf dx}_1}{|{\bf dx}_4 \times {\bf dx}_1|}
	\right)
\end{equation}
and (comparing Eqs.~(\ref{stokes}-\ref{stokesarea})) the current in this direction as
\begin{equation}
J_n^{(1)} = {\mathcal{I}}/{A}.
\end{equation}
In the same way, we can define $J_n^{(2)}$ and $J_n^{(3)}$ perpendicular to the other two mesh surfaces, with normal vectors $\nn^{(2)}$ and $\nn^{(3)}$, that pass through $X_{i,j,k}$. 

Now, the $J_n^{(p)}$ are projections of the current we require (denoted $\JJ_s$) in such a way that
\begin{equation}\label{normaleq}
\nn^{(p)}\cdot \JJ_s = J_n^{(p)}, \qquad p=1,2,3.
\end{equation}
Denoting by $\mathcal{N}$ the matrix whose rows are the row vectors $\nn^{(1)}$, $\nn^{(2)}$ and $\nn^{(3)}$, and by $\JJ_n$ the vector with components $J_n^{(1)},J_n^{(2)},J_n^{(3)}$ we can re-write Eq.~(\ref{normaleq}) as $\mathcal{N} \JJ_s = \JJ_n.$
Finally, we obtain the current at point $X_{i,j,k}$ via
\begin{equation}
\JJ_s = \mathcal{N}^{-1} \JJ_n.
\end{equation}
Note that $\mathcal{N}$ is always invertible assuming that the $\nn^{(p)}$ are linearly independent, i.e.~as long as the grid cells have non-zero volume. 
It is straightforward to verify that this procedure reduces to the standard 2nd-order centred difference expression in each direction for a rectangular (undeformed) mesh. 

Since this method makes use of Stokes' theorem, which is ``topological" in the sense that is does not depend on a deformation of the loop we are integrating over, we expect the method to be more robust and accurate for deformed grids \citep[see also][]{hyman1997}.
The algorithm has not yet been implemented in the full numerical scheme, as it requires a complete re-writing of the implicit time-stepping routine, and a simple explicit implementation turns out to be prohibitively slow to run at reasonable resolution. However, the algorithm is tested and used as a diagnostic in what follows.

\section{Comparison of methods}\label{ressec}
We now return to the test problem with two equal and opposite twists centred on $z=\pm z_0$ described above.

\subsection{Relaxing towards a rectangular mesh}
First, to demonstrate that in principle the original (2nd-order) relaxation scheme is sound, we perform the relaxation `experiment' in the following way. We begin with a uniform field ($\BB=b_0 \zhat$) on a uniform mesh, and then {\it deform this mesh} in such a way that the resulting magnetic field is $T2$ (with $\phi_i=\pm\pi$). This configuration is then relaxed, to a level where the Lorentz force calculated by the numerical scheme is reduced to $\JJ\times\BB=10^{-6}$. The equilibrium field corresponding to $T2$ is the uniform field $\BB=b_0\zhat$, and in this case (due to the frozen-in condition) the straight field should correspond to the rectangular mesh. (The choice $|\phi|=\pi$ is motivated by the braiding example discussed in \cite{wilmotsmith2009} since this is the minimum level of twist which yields a magnetic field whose field lines are truly braided in that case.)

To diagnose the success of the relaxation, referring back to Eq.~(\ref{fffquality}), we calculate
\begin{equation}\label{epsilonstar}
\epsilon^* = d \, \frac{\Delta\alpha^*}{\Delta l}
\end{equation}
taking $d=\sqrt{2}$ (radius of twist regions) and $\Delta\alpha^*$ and $\Delta l$ as the maximum change in $\alpha^*$ over a given length, which occurs along the central field line -- the $z$-axis -- by symmetry. This expression puts a true value on the `quality' $\epsilon^*$ of the force-free approximation: for a given variation in $\alpha^*$, $\epsilon^*$ provides a lower bound for the maximum value of $|\JJ\times\BB|/|\BB|^2$ within our domain, for a current free of errors (i.e.~setting $\delta=0$ in Eq.~(\ref{fffquality})). We obtain a value of $\epsilon^* = 9.5\times 10^{-7}$ for resolution $N=61$, demonstrating that {discretisation} errors in the scheme are very small when the relaxed state has an approximately rectangular mesh.

\subsection{Accuracy test: artificially imposed deformation}
We now investigate the promise of the two extensions to the scheme described in the previous section.
In our test case ($T2$) the topology of the field is simple, and the equilibrium field known, permitting the above approach (i.e.~relaxation towards a uniform mesh). However, to investigate magnetic fields with non-trivial topology we must approach the problem in a different way. We therefore return to the case where we begin with $T2$ on a rectangular mesh, setting $\phi_i=\pm\pi$. 

Performing the artificially imposed analytical (`untwisting') deformation  {\it instead of relaxation}, we see that derivatives calculated with the two new methods (4th-order finite differences and the Stokes-based method) both give smaller maximum errors than with the original 2nd-order scheme (see Table \ref{comptab}). {For a less deformed mesh ($\psi=\pi/2$), the 4th-order finite differences perform better than the Stokes-based routine. However, for a more distorted mesh ($\psi=\pi$), the Stokes routine gives significantly lower errors than either of the finite-difference methods. It is particularly interesting to note that the errors for the Stokes-based method seem to scale relatively weakly with the mesh deformation, suggesting it to be a good choice for highly deformed meshes. Both new schemes, for reasonable resolution and levels of deformation ($N\geq 41$, $\psi=\pi$) give errors that are an order of magnitude lower than the original (2nd-order) method.} This suggests these methods are worth pursuing, so we go on to perform relaxation simulations using them.

\begin{table}
\centering
\begin{tabular}{c || c|c|c |@{\hspace{0.2cm}}| c|c|c}
 & \multicolumn{3}{c|@{\hspace{0.2cm}}|}{$\psi=\pi$} & \multicolumn{3}{c}{$\psi=\pi/2$} \\ \cline{2-7}
$N$ & 2 n.n. & 4 n.n. & Stokes & 2 n.n. & 4 n.n. & Stokes\\ \hline&&&&&& \\[-1.4em] \hline 
21 & 670 & 222 & 22.1 & 37.8 & 12.2 & 18.7  \\ \hline 
41 & 144 & 19.4 & 5.85 & 10.6 & 1.01 & 4.71 \\ \hline 
61 & 64.7 & 4.14 & 2.69 & 4.99 & 0.551& 2.13 \\ \hline 
81 & 37.0 & 1.61 & 1.57 & 2.83 & 0.898 & 1.20
\end{tabular}
\caption{Errors in ${\bf J}$ for deformations with $\psi=\pi$ and $\pi/2$ (in Eq.~(\ref{analytangle})) using 2nd- and 4th-order finite differences (2 n.n./4 n.n., respectively) and the Stokes-based routine. $N^3$ is the mesh resolution. In each case the value shown is the maximum relative percentage error in the domain over all components of ${\bf J}$, i.e.  $100\times |J_i - J_i^a |_{max} / |J_i^a |_{max}$.}
\label{comptab}
\end{table}

\subsection{Relaxation with initially rectangular mesh}
We now leave the artificially imposed deformation, and relax $T2$ using both the 2nd- and 4th-order schemes. The relaxation is allowed to run until $|\JJ\times\BB|$ as calculated by the relevant numerical scheme is reduced to $10^{-5}$. We compare this value with $\epsilon^*$ (defined by Eq.~(\ref{epsilonstar})), and also the maximum value of the Lorentz force obtained by calculating $\JJ$ via the Stokes-based routine, denoted $\JJ_s$. 

\begin{figure}[t]
\centering
(a) \includegraphics[scale=0.5]{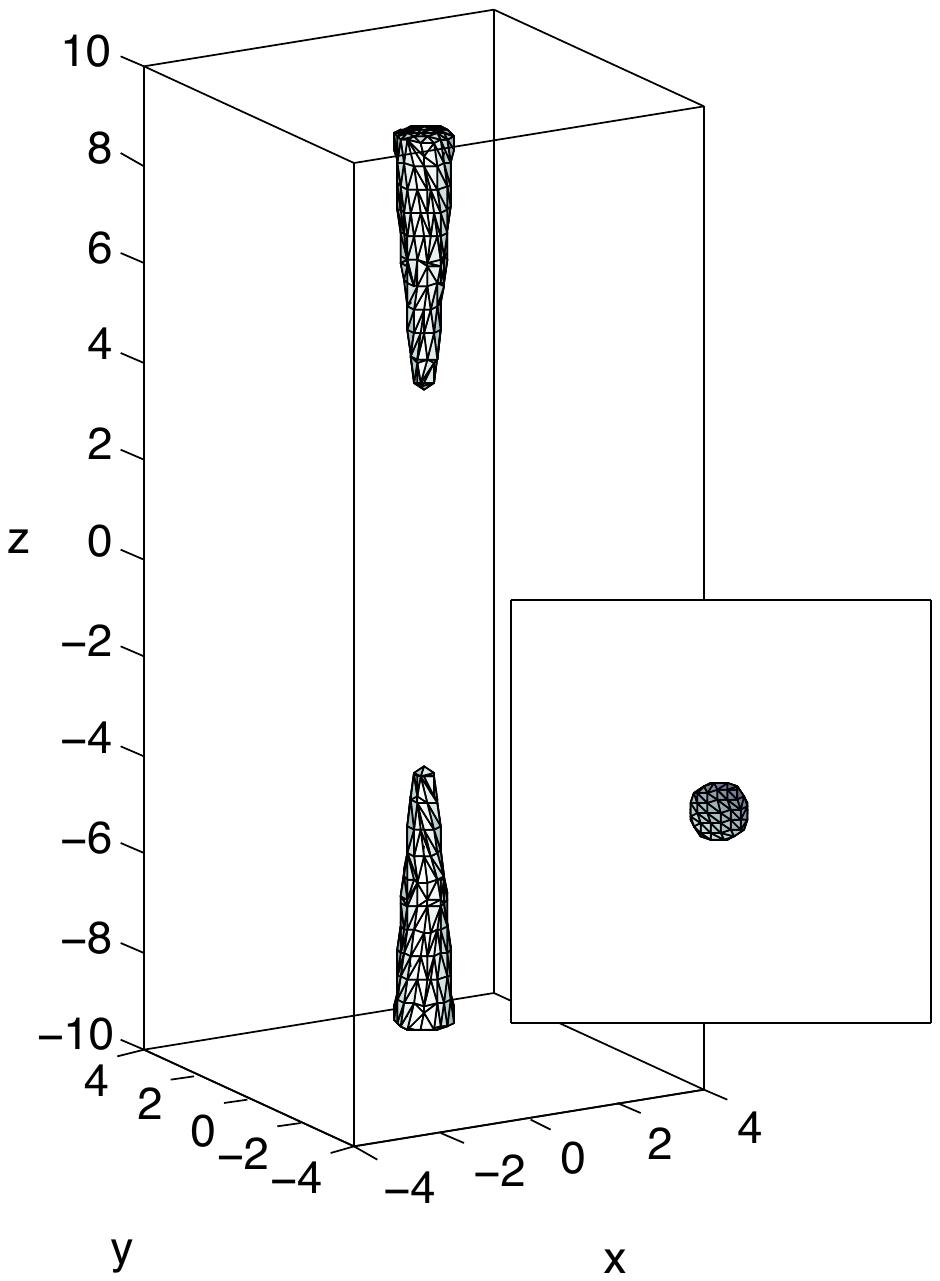}
(b) \includegraphics[scale=0.5]{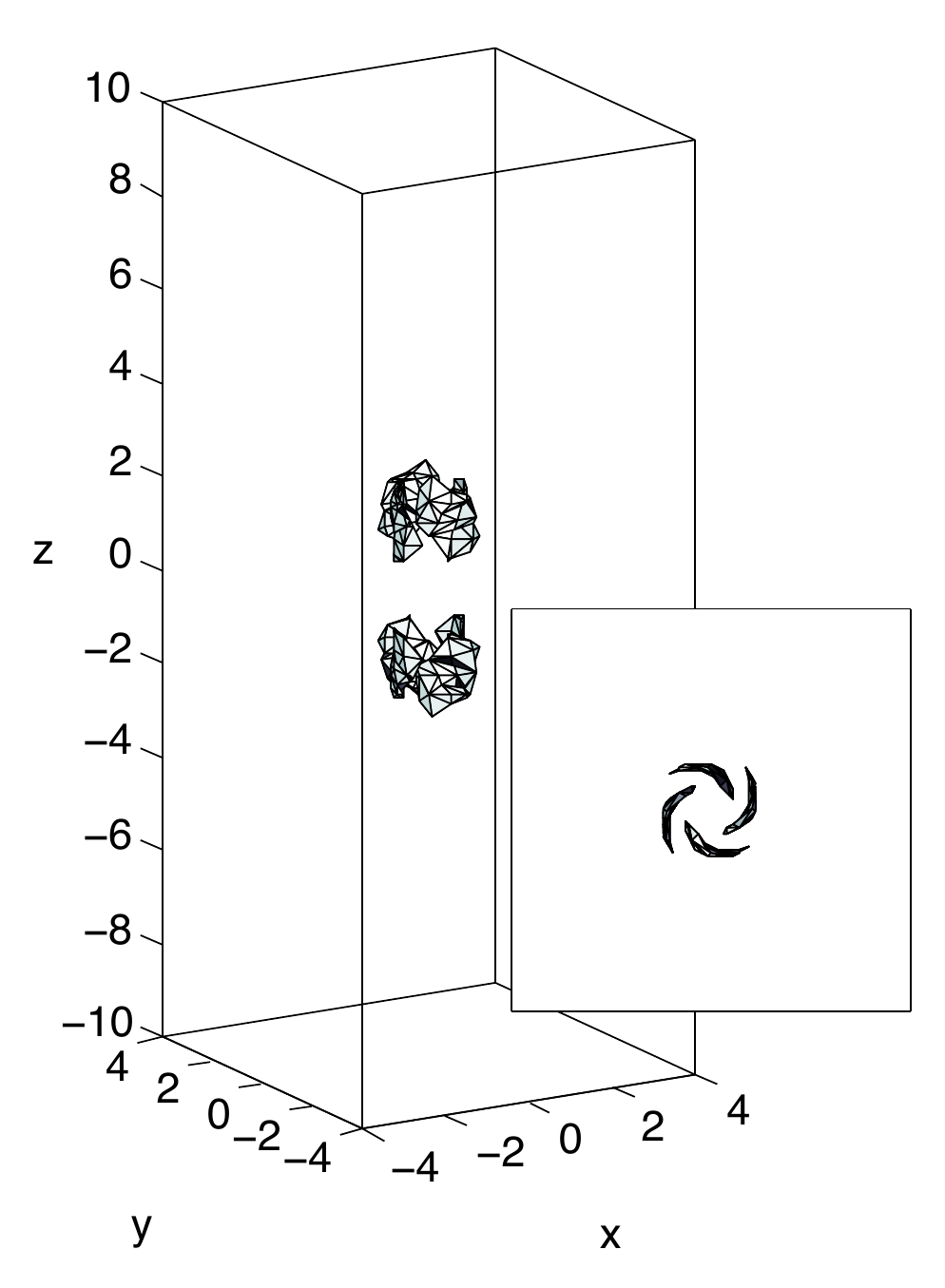}
\caption{Isosurfaces of $|\JJ|$ at 2/3 of maximum, for (a) an intermediate stage in the relaxation ($\JJ\times\BB=0.05$) and (b) the final state ($\JJ\times\BB=10^{-5}$). 4th-order scheme, $N=81$, $\phi_i=\pm\pi$. Inset: view in $xy$-plane (i.e.~from $z>10$).}
\label{jiso}
\end{figure}

The results for two levels of initial twist ($\phi_i=\pm\frac{\pi}{2},~ \phi_i=\pm\pi$) are displayed in Tables~\ref{tab2} and \ref{tab3}.
\begin{table}
\centering
\begin{tabular}
{c || c|c |@{\hspace{0.2cm}}| c | c}
 & \multicolumn{2}{c|@{\hspace{0.2cm}}|}{2 n.n.} & \multicolumn{2}{c}{4 n.n.}  \\ \cline{2-5}
$N$ & $\epsilon^*$ & $\JJ_s\times\BB/|\BB|^2$ &$\epsilon^*$ & $\JJ_s\times\BB/|\BB|^2$ \\ \hline&&&& \\[-1.4em] \hline 
21 & 0.11  & 0.053 & 0.048 & 0.063  \\ 
41 & 0.054  & 0.046 & 0.0067 &  0.018  \\ 
61 & 0.032 & 0.035 & 0.0042 & 0.0050  \\ 
81 & 0.021  & 0.027 & 0.0015 &  0.0019
\end{tabular}
\caption{Values of the force-free quality parameter $\epsilon^*$ and the Lorentz force calculated via the Stokes-based routine, $\JJ_s\times\BB/|\BB|^2$, for simulation runs with 2nd- and 4th-order finite differences (2 n.n.~/ 4 n.n.) and resolution $N^3$, to two significant figures. Twist parameter $\phi_i=\pm\pi/2$.}
\label{tab2}
\end{table}
\begin{table}
\centering
\begin{tabular}
{c || c|c |@{\hspace{0.2cm}}| c | c}
 & \multicolumn{2}{c|@{\hspace{0.2cm}}|}{2 n.n.} & \multicolumn{2}{c}{4 n.n.}  \\ \cline{2-5}
$N$ & $\epsilon^*$ & $\JJ_s\times\BB/|\BB|^2$ &$\epsilon^*$ & $\JJ_s\times\BB/|\BB|^2$ \\ \hline&&&& \\[-1.4em] \hline 
21 & 0.28  & 0.18 & 0.17 & 0.21  \\ 
41 & 0.16  & 0.17 & 0.071 &  0.13  \\ 
61 & 0.11 & 0.13 & 0.026 & 0.062  \\ 
81 & 0.074  & 0.11 & 0.021 &  0.023
\end{tabular}
\caption{As Table \ref{tab2}, with twist parameter $\phi_i=\pm\pi$.}
\label{tab3}
\end{table}
A number of points are immediately clear. First, in no simulation do we approach the apparent value of $\epsilon=10^{-5}$. However, the use of fourth- rather than second-order finite differences improves the quality of the relaxed field by an order of magnitude in $\epsilon^*$ when $|\phi|=\pi/2$. Increasing the deformation in the final state (by increasing $|\phi|$ in the initial state $T2$) has a strong adverse effect on the relaxation process. This is found to be because spurious (unphysical) current concentrations arise where none should reasonably be expected. Examining the corresponding mesh, we find that these `false' current regions appear where the grid is most distorted -- see Fig.~\ref{jiso}, and compare with Fig.~\ref{twoblobfig}(b).
The `current shards'  shown in Fig.~\ref{jiso}(b) actually intensify as the relaxation proceeds. We find that this is possible since $\nabla \cdot {\bf J}$ (approximated by interpolating $\JJ$ onto a uniform mesh) is not close to zero. As a result there is no `return current' associated with these localised current regions, which might be expected to generate a Lorentz force that would act against the further intensification of the current shards (if they have no physical basis). 
In previous studies using such codes \citep[e.g.][]{craiglitvinenko2005, pontincraig2005}, intensification of $|\JJ|$ as $|\JJ\times\BB|$ decreased in time was associated with current singularities, so at first sight it appears that these current shards could naively be interpreted as `current sheets', which would of course be unphysical. Note, however, that the most important signature of current singularity in previous studies was a (power-law) proportionality of the peak current with mesh resolution (for given $\epsilon$). We have found that in fact the current shards become less intense as $N$ is increased, so there is a clear distinction between the two phenomena.

\begin{figure}[t]
\centering
\includegraphics[scale=0.6]{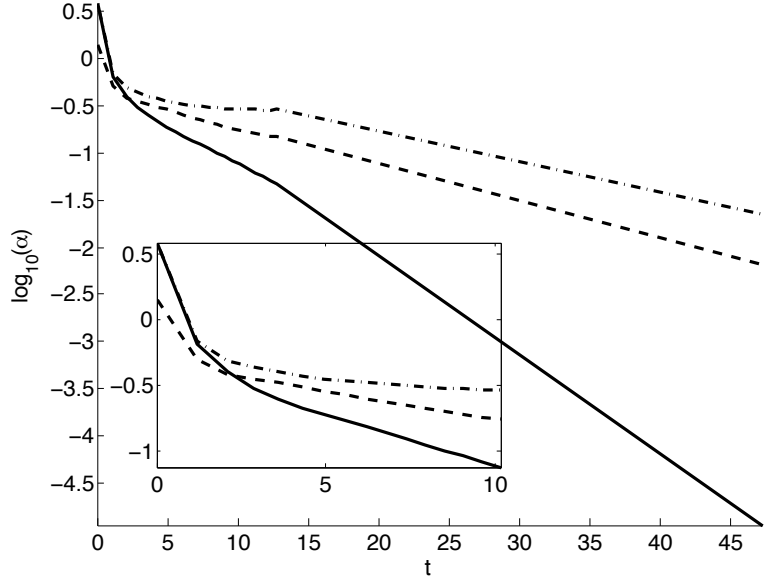}
\caption{Evolution of different $\alpha^*$'s through the relaxation with second order finite differences with $N=81$ and $\phi_i=\pm\pi/2$. The dashed line is the observed value of $\alpha^*$, while the solid line is the maximum allowable $\alpha^*$ defined via Eq.~(\ref{alphamax}) with $\epsilon$ given by $\JJ\times\BB/|\BB|^2$ from the 2nd-order numerical scheme. The dot-dashed line is the maximum $\alpha^*$ with $\epsilon$ given by $\JJ_s\times\BB/|\BB|^2$. Inset: close-up of behaviour at early time.}
 \label{alpha_t}
\end{figure}

Finally, consider the values of $\JJ_s\times\BB/|\BB|^2$ we find for the relaxed fields (Tables \ref{tab2}--\ref{tab3}). They are clearly of the same order as $\epsilon^*$ (note that $\epsilon^*$  based on $\JJ$ from the numerical scheme and $\epsilon^*$ based on $\JJ_s$ are of the same order), and thus it seems that an implementation involving the Stokes-based routine has the capacity to yield a magnetic field that is much closer to being force-free  (with lower $\epsilon^*$). This is illustrated in Fig.~\ref{alpha_t}.
We consider the observed value of $\alpha^*$ based on Eq.~(\ref{alphastar}) (dashed line), compared with a maximum allowable value for $\alpha^*$. Since $\alpha^*$ is anti-symmetric about $z=0$, the maximum value allowed, based on Eq.~(\ref{fffquality}), is
\begin{equation}\label{alphamax}
\alpha^*_{max} = \frac{L \epsilon}{d},
\end{equation}
and we take $|\BB|=1$ and $d=\sqrt{2}$ as before, and $L=20$, the length of the domain. Then we obtain the maximum possible $\alpha^*$ by taking $\epsilon$ to be the maximum value  during the relaxation of $\JJ\times\BB$ (solid line) or $\JJ_s\times\BB$ (dot-dashed). We see that very early in the relaxation the actual value of $\alpha^*$ becomes greater than the maximum allowed by $\JJ\times\BB$ from the numerical scheme (2nd-order). Moreover, the discrepancy grow steadily. However, $\alpha^*$ always remains less than the maximum allowed by the Stokes-based method, implying that this may be a more sound method to calculate the current and resulting Lorentz force.

\section{Conclusions}\label{concsec}
Force-free magnetic fields are important in many astrophysical applications.
Determining the properties of such force-free fields -- especially smoothness and stability properties -- is key to understanding energy release processes that heat the plasma and lead to dynamic events such as flares in the solar corona.
We have investigated the properties of different relaxation procedures for determining force-free fields based on a Lagrangian mesh approach. These techniques have previously been shown to have many powerful and advantageous properties. Previous understanding was that such schemes would iteratively converge (i.e.~$\JJ\times\BB$ decreasing monotonically to a given level) up to a certain degree of mesh deformation. Beyond this level of mesh deformation the scheme no longer converges ($\JJ\times\BB$ oscillates or grows), and it is this phenomenon that was thought to limit the method. 
However, we have shown above that even when the numerical scheme iteratively converges, the accuracy of the force-free approximation can become seriously compromised for even `moderate' mesh deformations. This error is an accumulation of {numerical discretisation} errors resulting from the calculation of $\JJ$ via combinations of 1st and 2nd derivatives of the mesh deformation Jacobian -- which are calculated using finite differences. The result is that neither $\JJ=\nabla \times {\bf B}$ nor subsequently $\nabla \cdot {\bf J}=0$ are well satisfied.

{
It was demonstrated that a result of the breaking of the solenoidal condition for $\JJ$ can be the development of spurious (unphysical) current structures. However, we note that these rogue currents do diminish with resolution ($N$), so when using these schemes this property should always be checked where possible.  We expect that, as a result, if it were possible to systematically increase $N$ indefinitely the rogue currents would eventually vanish. In other words, the real problem is that the iterative convergence (i.e.~monotonic decrease of $|\JJ\times\BB|$ with $t$) is not compromised by the rogue currents but the real convergence to a correct solution is severely impaired.
}

A force-free field is defined by $\nabla \times {\bf B}=\alpha\BB$. One key result of this equation is that $\alpha$ must be constant along magnetic field lines. We therefore argued that a correct diagnostic  to measure the quality of a force-free apprioximation is the constancy of the parameter $\alpha^*=J_{\|}/|\BB|$ along field lines. An appropriate normalisation is given in Eq.~(\ref{epsilonstar}). {The results of our investigations suggest that for Lagrangian schemes the $|\JJ\times\BB|$ measure does not provide  a  good indicator of true convergence----a better measure in the constancy of $\alpha^*$ along $\BB$.
} We note that other authors have proposed measures other than the maximum of $\JJ\times\BB$ for testing a force-free approximation -- for example \cite{wheatland2000} introduced the ``mean current-weighted angle between $\JJ$ and $\BB$". However, calculation of this measures still relies upon the value of $\JJ\times\BB$ in the numerical scheme, and in the present scenario we have shown that the errors arise not because $\JJ$ and $\BB$ are not parallel, but because $\nabla\cdot\JJ\neq 0$.

Since errors in the (Lagrangian) numerical scheme investigated here arise as the mesh becomes increasingly distorted, a natural choice is to begin with a non-equilibrium field on a non-rectangular mesh, and relax towards a (perhaps approximately) rectangular one. However, this approach  is not feasible if the field has complex topology. In the case of a braided field -- which is of particular interest to the theory of the solar corona -- we find that for our realisation of such a field \citep{wilmotsmith2009} there is no escaping having at least a moderately distorted mesh in the final state. 

We proposed two possible extensions to the numerical method. The first was to increase the order of the finite differences used. It was found that for certain levels of deformation this can give an order of magnitude improvement in the quality of the force-free approximation obtained. It is therefore certainly a good approach to use in some circumstances. 
As the mesh became more and more highly deformed, the advantage of the scheme with 4th-order finite differences was lost for our test case $T2$. Furthermore, we found that for relaxation of the braided field described in \cite{wilmotsmith2009} no appreciable improvement arose from using the 4th-order scheme.

The other extension that we proposed to the scheme seems very promising. In Section \ref{stokessec} we presented an algorithm for calculating the curl of a vector field on an arbitrary mesh, based on Stokes' theorem. { For increasing levels of mesh deformation, this performed progressively better  than the finite difference methods}. What's more, in all of our tests the resultant Lorentz force $\JJ_s\times\BB$ had lower errors than that calculated by the traditional finite difference. In order for a relaxation experiment to remain accurate as it proceeds, the maximum allowed value of $\alpha^*$ based on $\JJ\times\BB$ (see Eq.~(\ref{alphamax})) must always remain greater than the maximum observed value of $\alpha^*$. We found that this is the case for the Stokes-based $\alpha^*$ down to at least an order of magnitude lower in $\JJ_s\times\BB$ than for the finite difference methods (see Fig.~\ref{alpha_t}).

All of the above leads us to believe that the Stokes-based algorithm is a highly promising one for improving the accuracy of Lagrangian relaxation schemes. At present it has not been implemented (i.e.~the code does not act to minimise $\JJ_s\times\BB$) because this requires a complete re-writing of the implicit (ADI) time-stepping, and a simple explicit implementation turns out to be prohibitively computationally expensive. However, our intended next step in this investigation is to implement this scheme, either by introducing the Stokes-based current calculation as a correction term in the existing scheme or by employing a more sophisticated explicit time-stepping to reduce the computational expense to acceptable levels. 
We note that while the algorithm at present only uses two nearest neighbour points in each direction, it could be extended to include further line integrals as corrections to the present formula for $\JJ_s$ in much the same way as is done by increasing the order of finite difference derivatives.


\acknowledgments

The authors are grateful to A.~Nordlund and K.~Galsgaard for
helpful discussions.




\end{document}